\documentclass[12pt]{article}
\usepackage{fullpage,epsfig,graphics,amsbsy,amssymb,cancel}
\usepackage{psfrag,hyperref}
\usepackage{graphicx,color}
\newcommand{\be}{\begin{eqnarray}}
\newcommand{\ee}{\end{eqnarray}}
\usepackage{amsmath}
\numberwithin{equation}{section} 
\begin{document}

\thispagestyle{empty}
\begin{flushright} \small
UUITP-04/12
 \end{flushright}
\smallskip
\begin{center} \LARGE
{\bf  Twisted supersymmetric 5D Yang-Mills theory and contact geometry}
 \\[12mm] \normalsize
{\bf  Johan~K\"all\'en and Maxim Zabzine} \\[8mm]
 {\small\it
  Department of Physics and Astronomy, 
     Uppsala university,\\
     Box 516, 
     SE-75120 Uppsala,
     Sweden
   }
\end{center}
\vspace{7mm}
\begin{abstract}
 \noindent

  We extend the localization calculation of the 3D Chern-Simons partition function over Seifert manifolds to an analogous calculation 
   in five dimensions. We construct a twisted version of $N=1$ supersymmetric Yang-Mills theory defined on a circle bundle over a four dimensional
     symplectic manifold.  The notion of contact geometry plays a crucial role in the construction
       and we suggest a generalization of the instanton equations to five-dimensional contact manifolds. 
   Our main result is a calculation of the full perturbative partition function on $S^5$
    for the twisted supersymmetric Yang-Mills theory with  different Chern-Simons couplings. The final answer is given in terms of a
     matrix model. Our construction admits generalizations to higher dimensional contact manifolds. This work is inspired by the work of Baulieu-Losev-Nekrasov from the mid 90's, and in a way it is covariantization of their ideas for a
  contact manifold. 
 \end{abstract}

\eject
\normalsize
 
\section{Introduction} 
 
 In  \cite{Kallen:2011ny}, a twisted version of 3D $N=2$ supersymmetric Chern-Simons theory was defined for any contact three manifold $M_3$.  
  Any compact, orientable three manifold admits  a contact structure and thus $N=2$ twisted Chern-Simons theory can be defined for 
   any such manifold. This theory is a non-dynamical extension of the standard Chern-Simons theory. In  the case of a Seifert manifold ($U(1)$-bundles over Riemann surfaces), 
    the partition function of this theory can be calculated exactly using localization techniques and it coincides with the known results.
   
         On Seifert manifolds the BRST-exact Lagrangian of the theory is the action for  twisted 3D $N=2$ supersymmetric Yang-Mills theory. Thus the partition function for 
      Chern-Simons theory on a Seifert manifold can be interpreted in two ways, either as the partition function of Chern-Simons theory or 
       as the partition function of $N=2$ twisted supersymmetric Yang-Mills. It turns out that this story has a rather straightforward 
        five-dimensional generalization, and this is the subject of this paper. Our present work is inspired by the work of Baulieu-Losev-Nekrasov \cite{Baulieu:1997nj} and an earlier paper by Nekrasov \cite{Nekrasov:1996cz}.
  In this work we concentrate on the 5D theory, although many results have a straightforward generalization to any odd dimensional manifold
   admitting a contact structure. 
 
 Let us briefly sketch the main idea. Consider a  $d$-dimensional manifold $M_d$ with a $G$-bundle over it. Let $A$ be a connection one-form on $M_d$. Introduce other fields; $\Psi$ is an odd (fermonic)  one-form transforming in the adjoint representation, $\sigma$ is a real scalar, also in the adjoint. Finally $\chi^\alpha$ is collection of odd (fermonic)  differential forms with possible further restriction on them 
   and $H^\alpha$ is collection of even (bosonic) differential forms with possible restriction on them, both in the adjoint representation. If we pick   
a vector field $v$ on $M_d$ then we can introduce
 the following complex transformations
 \begin{equation} \label{general-transform}
 \begin{split}
 &\delta A = \Psi~, \\
 & \delta \Psi = \iota_v F + i d_A \sigma~, \\
 & \delta \sigma = - i~ \iota_v \Psi~,\\
 & \delta \chi^\alpha = H^\alpha~,  \\
 & \delta H^\alpha =  {\cal L}^A_v \chi^\alpha - i [\sigma, \chi^\alpha]~, 
 \end{split}
 \end{equation}  
  where we use the following notations:  $d_A = d + [A, ~]$ is the de Rham operator twisted by the connection, ${\cal L}_v^A = {\cal L}_v + [\iota_v A, ~]$ with 
   $\mathcal{L}_v$ being  the Lie derivative along $v$ and $F= dA + A \wedge A$ is the curvature two-form for the connection $A$.
    These transformations are defined in any dimension with a choice of vector field $v$. They satisfy the following algebra
     \be
      \delta^2 = {\cal L}_v + G_\Phi~, 
    \ee
    where $G_\Phi$ denotes a  gauge transformation with parameter $\Phi = i \sigma - \iota_v A$ and it is defined as 
     follows      
\be
G_\Phi A = d_A \Phi~,~~~~~G_\Phi X= [ X, \Phi]~,
 \ee
 where $X$ stands for  all fields transforming in the adjoint representation. With these transformations at hand, the natural question is which observables, if any,
are invariant under them. It turns out that the construction of observables requires more structure on 
   $M_d$. We will show that if $M_d$ is an odd dimensional contact manifold then we can always construct a supersymmetric extension  
    of different Chern-Simons terms. 
      In addition, if $M_d$ is a $U(1)$-bundle over a symplectic manifold (this is a special case of a contact manifold),
       then the twisted supersymmetric version of 5D Yang-Mills theory can be chosen as the Lagrangian for the theory. The fields in this theory can be mapped to the field content of the standard supersymmetric 
  $N=1$ Yang-Mils theory in five dimensions, and our theory can be thus be thought of as a twisted version of that theory. Also, in this case the partition function 
       can be calculated using localization techniques. 
 
As an example of an application of the localization technique, we present the calculation of the full perturbative partition function on $S^5$, for the twisted supersymmetric Yang-Mills theory coupled to two different Chern-Simons terms. 
       The result of the calculation is given by an interesting matrix model.   Our result leads to an immediate puzzle since, due to standard arguments,   $N=1$ supersymmetric Yang-Mills theory in five dimensions should be perturbatively nonrenormalizable, and at the same 
        time using the localization technique we are able to compute the full perturbative answer. At the present moment, we are unable to suggest 
         an explanation of this puzzle.   However, see the recent works \cite{Lambert:2010iw, Douglas:2010iu} regarding the UV-completion of 5D maximal supersymmetric Yang-Mills theory.
         
 The paper is organized as follows. In section \ref{3D} we review a setup of for doing localization of 3D Chern-Simons theory. We mainly follow
  the work \cite{Kallen:2011ny}, although we present some new observations and try to put the paper \cite{Kallen:2011ny}
   in a wider context. In section \ref{5D-theory-const} we construct the 5D analog of this story. We discuss supersymmetric Chern-Simons 
    terms, the five-dimensional analog of instantons and the construction of the twisted supersymmetric 5D Yang-Mills theory. We also explain the necessary geometrical ingredient needed for the construction to work. 
      In  section \ref{calculation} we perform the perturbativly exact calculation of the partition 
       function on $S^5$. The result is given by a matrix model. This calculation admits 
        numerous generalization which we briefly discuss in section \ref{end}. 
        Section \ref{end} also gives a summary of the results and future outlook.  In order to make the paper more readable we have 
         collected many technicalities in appendices. Appendix \ref{geometry-cont} gives a short summary of the contact geometry we need in this paper and fixes 
         different normalizations. In Appendix \ref{GCS-terms} we present generalization of supersymmetric Chern-Simons terms to 
         higher dimensional contact manifolds. Appendix \ref{1loopcalc} collects the technical points related to the calculation on $S^5$.

 \section{3D gauge theories and contact structures}
 \label{3D}

In this section we review the twisting of 3D $N=2$ supersymmetry and its relation to contact geometry in order to set the stage for higher dimensional generalizations. We will also discuss the relation between the different approaches to 3D Chern-Simons theory that has appeared in \cite{Kallen:2011ny}, \cite{Beasley:2005vf} and \cite{Thompson:2010iy}.

\subsection{3D Chern-Simons theory and twisted supersymmetry}

Let $G$ be a compact, simple and simply connected Lie group. Let us consider the standard Chern-Simons theory
  \begin{equation}
  S_{CS_3} (A) = \frac{\rm k}{4\pi} \int\limits_{M_3} {\rm Tr}  \left ( A \wedge d A + \frac{2}{3} A \wedge A \wedge A \right )~,\label{classical-CS}
  \end{equation}
   which is defined for a trivial $G$-bundle over $M_3$. $A$ is the connection one-form, and ${\rm Tr}$ denotes an invariant inner product on $\mathfrak{g}$, the Lie
     algebra of $G$. In order for the quantum theory to be gauge invariant we must impose that the level $\rm k$ is an integer; ${\rm k} \in \mathbb{Z}$. 

 Let us on $M_3$ choose a contact structure, which can be defined by a contact form $\kappa \neq 0$, such that $\kappa \wedge d \kappa \neq0$. We can always define a vector field $v$, known as the Reeb vector field, with the property $\iota_v \kappa = 1$ and $\iota_v d \kappa=0$. We can define the action
\begin{equation}
 S_{SCS_3} =   \frac{\rm k}{4\pi}  \int\limits_{M_3}  {\rm Tr} \left ( A \wedge d A + \frac{2}{3} A \wedge A \wedge A  - \kappa \wedge \Psi \wedge \Psi - 
  2 \kappa\wedge d\kappa 
   ~\iota_v\Psi \alpha + \kappa \wedge d\kappa D \sigma \right )~,\label{main-action}
\end{equation}
where $\Psi$ is an odd one-form, $\alpha$ is odd scalar, $D$ and $\sigma$ are even scalars, and all these  fields are in the adjoint representation of the gauge group. The exponent of this action is gauge invariant. Moreover, the action (\ref{main-action}) is invariant under the following complex BRST symmetries
  \begin{equation}\label{3D-twisted-tran}
  \begin{split}
 &\delta A = \Psi~, \\
 &\delta \Psi = \iota_v F + i d_A \sigma~,\\
  &\delta \sigma = - i~ \iota_v \Psi~, \\
 & \delta  \alpha  =   - \frac{\kappa\wedge  F}{\kappa \wedge d\kappa} + \frac{i}{2} (\sigma + D)~,\\
 & \delta D = - 2 i {\cal L}_v^A \alpha - 2[\sigma, \alpha]
 -2i \frac{\kappa \wedge   d_A \Psi}{\kappa \wedge d\kappa} +i ~ \iota_v \Psi ~.
 \end{split}
 \end{equation}

 Originally in \cite{Kallen:2011ny}, the action (\ref{main-action}) and the  transformations (\ref{3D-twisted-tran}) were motivated and obtained 
  from a twisting of the supersymmetry transformations on $S^3$ discussed
   in \cite{Kapustin:2009kz}. However, written in the present form, the action and transformations are defined for any contact manifold $M_3$.
    Indeed, any three manifold admits a contact structure and thus this cohomological action can defined for any $M_3$ upon
     the choice of a contact structure.  Another observation is that the transformations  (\ref{3D-twisted-tran})
   are the same as in (\ref{general-transform}) upon the following field redefinitions
 $$\chi= \alpha~,~~~~~~~H = - \frac{\kappa \wedge F}{\kappa \wedge d\kappa} +\frac{i}{2} (\sigma +D)~,$$
  where now $\chi$ is a fermionic zero-form  and $H$ is a bosonic  zero-form.

\subsection{A different version of 3D Chern-Simons theory} 

The cohomological form (\ref{main-action}) of the Chern-Simons action presented in the previous subsection can be reformulated 
 in different ways. Let us add the BRST-exact term  
  \begin{equation}
   S^{new}_{SCS_3} = S_{SCS_3} +  \frac{\rm k}{4\pi}  \int\limits_{M_3} {\rm Tr} \left ( \delta (2i \alpha \sigma ~\kappa \wedge d\kappa) \right ) \label{new-action}
  \end{equation}
   to the action (\ref{main-action}). The addition of this term does not change anything since 
   \be
  \int\limits_{M_3} {\rm Tr} \left ( \delta^2 (2i \alpha \sigma ~\kappa \wedge d\kappa ) \right ) = 0
   \ee
   and the partition function remains the same due to standard localization arguments. 
   The new action (\ref{new-action}) has the form
   \begin{equation}
    S^{new}_{SCS_3} =  \frac{\rm k}{4\pi} \int\limits_{M_3} {\rm Tr} \left (   A \wedge d A + \frac{2}{3} A \wedge A \wedge A  - \kappa \wedge \Psi \wedge \Psi  - 2i \kappa \wedge F \sigma -
     \sigma^2 \kappa \wedge d\kappa \right ) ~.\label{new-CS-shift}
   \end{equation}
 By construction,   this action is invariant under the BRST transformations (\ref{3D-twisted-tran}) involving only $A$, $\Psi$ and $\sigma$.  
  However, this action  has an additional gauge shift-symmetry
     \begin{equation} \label{shift}
     \tilde{\delta} A = \xi \kappa~,~~~~~~~~\tilde{\delta} \sigma = - i \xi~,
     \end{equation}
     where $\xi$ is an arbitrary function.
  This is manifest if we rewrite the  action $S^{new}_{SCS_3}$ as follows
  \begin{equation}
  S^{new}_{SCS_3} = S_{CS_3} (A - i\sigma \kappa) -  \frac{\rm k}{4\pi} \int\limits_{M_3} {\rm Tr} \left (\kappa \wedge \Psi \wedge \Psi \right )~. \label{shift-3D-action}
  \end{equation}
  The action (\ref{shift-3D-action}) is also invariant under the following symmetry
\begin{equation} \label{kresc}
\begin{split}
 A~\rightarrow~A~,~~~\kappa~\rightarrow~ e^f \kappa~,~~~\Psi~\rightarrow~e^{-f/2} \Psi~,~~~\sigma~\rightarrow~e^{-f}\sigma~, 
 \end{split}
\end{equation}
 where $f$ is a function on $M_3$. Thus the action is independent of the contact form and depends only on 
  the contact planes, ${\rm ker}(\kappa) \subset T M_3$.

 If we integrate out the non-dynamical field $\sigma$ we get the following action
    \be
    S^{new}_{SCS_3} =  \frac{\rm k}{4\pi} \int\limits_{M_3} {\rm Tr} \left (  A \wedge d A + \frac{2}{3} A \wedge A \wedge A  - \kappa \wedge \Psi \wedge \Psi  - \left ( \frac{\kappa\wedge F}{\kappa\wedge
     d\kappa} \right )^2 \kappa \wedge d\kappa \right ) ~,
    \ee
   which appeared previously in the discussion of non-abelian localization by Beasley and Witten \cite{Beasley:2005vf}. Moreover, the action (\ref{shift-3D-action}) (without the shift of the gauge field) has appeared in the work by Thompson \cite{Thompson:2010iy}.

We can therefore conclude that one interpretation of the relation between the approaches in \cite{Kallen:2011ny} and \cite{Beasley:2005vf} is that the action (\ref{main-action}) is the gauged fixed version 
  of the model (\ref{shift-3D-action}). We have introduced the additional fields $\alpha$ and $D$ to impose the condition $\kappa\wedge F=0$ to gauge fix the shift symmetry. 
      
\subsection{3D Yang-Mills theory and localization}      

Now let us return to the transformations (\ref{3D-twisted-tran}) and discuss further possible observables. 
Let us choose a metric $g$ on $M_3$ such that for a $p$-form $\omega$ we have
 $\iota_v * \omega = (-1)^p *( \kappa \wedge \omega )$. Such a metric always exists; see Appendix \ref{geometry-cont} for further explanations and conventions.

  We can write the following BRST exact term
  \begin{equation} \label{de3d}
 S_{SYM_3} =   \int\limits_{M_3}{\rm Tr}  \left ( \delta( \Psi \wedge * \overline{\delta \Psi}  + 4 \alpha \wedge * \overline{\delta \alpha})  \right )~,
  \end{equation}
   which is not automatically BRST-invariant since $\delta^2 \neq 0$.  However we can easily deduce that $\delta S_{SYM_3}=0$ if and 
    only if ${\cal L}_v g=0$.  Thus the metric should be invariant under the Reeb vector field. This situation is realized when the Reeb vector
     field $v$ corresponds to a $U(1)$-action on $M_3$ and thus when $M_3$ is a Seifert manifold. The explicit form of $S_{SYM_3}$ is given by
 \begin{equation} \label{3dym}
 \begin{split}
S_{SYM_3} =& \int\limits_{M_3} {\rm Tr} ( F \wedge *F + d_A \sigma \wedge * d_A \sigma  +  (\sigma +D)^2 +
  4 \alpha \wedge * ( {\cal L}^A_v \alpha - [i\sigma, \alpha])   \\
&   - \Psi \wedge * (\iota_v d_A \Psi - d_A \iota_v \Psi + [i\sigma, \Psi]) 
  -4 \alpha \kappa \wedge d_A \Psi  )~,
\end{split}
\end{equation}
 which is the twisted N=2 supersymmetric extension of 3D Yang-Mills theory. Thus, on a Seifert manifold, we can consider the theory 
  with action 
  \be 
   S_{SCS_3} + t S_{SYM_3}~,
  \ee
  where $t$ is the coupling for the supersymmetric Yang-Mills theory. Using the localization technique the partition function can be 
   calculated for this theory. The first term in $\eqref{de3d}$ imposes the conditions
   \be
   \iota_v F& =0~, \\
   d_{A}\sigma &=0~,
   \ee
whereas the second term imposes the condition
\be
\kappa\wedge F=0~.
\ee
Combining the two we get the conditions
\begin{equation} \label{3dmodspace}
\begin{split}
F&=0~, \\   
 d_A \sigma &= 0~.
\end{split}
\end{equation}
We notice that there is no reference to the contact structure in these equations, and it turns out that the partition function coincides with the partition function of the standard Chern-Simons theory on Seifert manifolds. For further details, see \cite{Kallen:2011ny}. 

Alternatively, we can start from the action (\ref{shift-3D-action}) and use the multiplet (\ref{general-transform}) to impose the gauge condition $\kappa \wedge F=0$:
\begin{equation}\label{3dymgf}
 S^{new}_{SCS_3}  + t  \int\limits_{M_3}{\rm Tr}~  \delta \left ( \Psi \wedge * \overline{\delta \Psi}  + 8 \chi \wedge * \left (\frac{\kappa \wedge F}{\kappa \wedge d\kappa} - \frac{H}{2} \right )  \right )~.
\end{equation}
 Again the BRST-exact term can be made BRST-invaraint on Seifert manifolds. Upon integration of the non-dynamical field $H$,
 this BRST-exact term is just the twisted supersymmetric Yang-Mills theory given by equation $\eqref{3dym}$ (without the $(\sigma+D)^2$-term). As before, it can be used for a localization calculation 
  of the Chern-Simons partition function and it produces the same answer.   

 These two actions and two approaches are essentially the same and they are related by some field redefinitions and some 
  BRST-exact terms. The first action $\eqref{3dym}$ and transformations $\eqref{3D-twisted-tran}$ have its origin in supersymmetry \cite{Kapustin:2009kz}, while the second action $\eqref{3dymgf}$ and transformations $\eqref{general-transform}$ are motivated by more direct topological field theory considerations, and they are generalization of the construction in \cite{Baulieu:1997nj} to contact manifolds. 

  As we have seen, there are many different ways of looking at the above three-dimensional theory. The logic of the approach which most directly generalize to higher dimensions can be described as follows. First, with a choice of contact structure, we can decompose the space of differential forms into a horizontal part and a vertical part:
\begin{equation}
\beta=\iota_v(\kappa\wedge \beta)+\kappa\wedge \iota_v\beta= \beta_H + \beta_V~, \quad \quad \beta\in\Omega^{\bullet}~.
\end{equation}
We introduce the transformations $\eqref{general-transform}$ in order to localize the theory to a finite dimensional space. The first three lines of $\eqref{general-transform}$ will impose the condition that the \textit{vertical} part of $F$ vanishes, together with the condition that the field $\sigma$ is covariantly constant. The $(\chi^{\alpha},H^{\alpha})$ multiplet is used to impose conditions on the \textit{horizontal} part of $F$. In general, we write these conditions as $G(F_H)=0$. In three dimensions, $G(F_H)=F_H$ and we only need one scalar multiplet $(\chi,H)$ to impose this condition. The Lagrangian of the theory is given by the second term in $\eqref{3dymgf}$, whereas the first term, the supersymmetric Chern-Simons functional $\eqref{shift-3D-action}$, is a non-trivial observable in this theory.

 \section{5D gauge theory and contact structures} \label{5D gauge theory}
 \label{5D-theory-const}
 
 In this section we will generalize the 3D construction to five dimensions. Let us consider a 5-dimensional contact manifold $M_5$ without boundary. It means that $M_5$ admits a one-form $\kappa \neq 0$
   such that $\kappa \wedge d\kappa \wedge d\kappa \neq 0$.  The Reeb vector field $v$ is defined as $\iota_v \kappa = 1$
     and $\iota_v d \kappa=0$. There are plenty of examples of five-dimensional contact manifolds, see for example Chapter 8 in \cite{MR2397738}. 

 Following the logic described in the last section, we start by discussing which conditions we will impose on the horizontal part of the curvature. We then proceed with describing which observables we can define in the five-dimensional theory. 
From now we always assume that  $M_5$  carries a contact structure and that it has no boundary.

\subsection{5D contact instantons}     
\label{gauge-fix-sub}
As mentioned above,  on a general  contact manifold we can introduce the analog of vertical and horizontal decomposition of differential forms
$\Omega^\bullet (M_{2n+1})$  by using the projectors $\kappa \wedge \iota_v$ and $(1-\kappa\wedge \iota_v)$. 
  Let us concentrate on a five-dimensional contact manifold $M_5$ and pick up a metric $g$ which has the 
   property $\iota_v ( * \omega_p) = (-1)^p * (\kappa \wedge \omega_p)$ for any $p$-form $\omega_p$. Such a metric 
    always exists.  Let us concentrate on two-forms $\Omega^2(M_5)$. Any two-form $\omega$ can be decomposed into 
     vertical and horizontal parts as follows
     \be
      \omega = (1 - \kappa \wedge \iota_v) \omega + \kappa \wedge \iota_v \omega = \omega_H + \omega_V~. 
     \ee
     Moreover, horizontal two-forms $\Omega_H^2(M_5)$ can be decomposed further as follows
     \be
      \omega_H = \frac{1}{2} ( 1+ \iota_v *) \omega_H +  \frac{1}{2} ( 1-  \iota_v *) \omega_H = \omega_H^+ + \omega_H^-~,
     \ee
     where $1/2 (1\pm \iota_v *)$ are projectors on $\Omega^2_H(M_5)$ (remember that in 5D, $*^2=1$ for all forms).  
      Thus two-forms on a contact manifold $M_5$ with compatible 
      metric can be decomposed as follows
    \be
     \Omega^2 (M_5) = \Omega^2_V (M_5) \oplus \Omega^{2+}_H (M_5) \oplus \Omega^{2-}_H (M_5)~.
    \ee 
   These three spaces are orthogonal to each other with respect to the standard scalar product on forms
   \be
( \alpha, \beta) = \int\limits_{M_5} \alpha \wedge * \beta~. 
   \ee 
  Let us now consider the curvature two-form $F$. The Yang-Mills action can be written as follows
   \begin{equation}\label{5D-YM-H-V}
    \int\limits_{M_5} {\rm Tr} ( F \wedge * F) =    \int\limits_{M_5} {\rm Tr} ( F_H \wedge * F_H + F_V \wedge *F_V)~,
   \end{equation}
    where $F_H = \iota_v (\kappa \wedge F)$ is the horizontal part of $F$ and $F_V = \kappa \wedge \iota_v F$ is the vertical part. Furthermore, 
     the horizontal part can be decomposed into $F^\pm_H = 1/2( F_H  \pm \iota_v (*F))$   and
     we have the following identities
   \begin{equation}
     \int\limits_{M_5} {\rm Tr} ( F_H^\pm \wedge * F_H^\pm) =    \frac{1}{2}\int\limits_{M_5} {\rm Tr} ( F_H \wedge * F_H )
     \pm \frac{1}{2} \int\limits_{M_5} {\rm Tr} (\kappa \wedge F \wedge F)~.\label{sq-self-dual-5D}
   \end{equation}
 Using the identity $a^2 + b^2 \geq |a^2 - b^2|$ and (\ref{sq-self-dual-5D}) we get the following bound
 \be
  \int\limits_{M_5} {\rm Tr} ( F_H \wedge * F_H ) \geq \left  | ~ \int\limits_{M_5} {\rm Tr} (\kappa \wedge F \wedge F)~ \right |~,
 \ee
 which can be combined further with (\ref{5D-YM-H-V}) to obtain the following bound on the 5D Yang-Mills action 
 \begin{equation}
  \int\limits_{M_5} {\rm Tr} ( F \wedge * F) \geq \left  | ~ \int\limits_{M_5} {\rm Tr} (\kappa \wedge F \wedge F)~ \right |~.\label{full-5D-bound}
 \end{equation}
  This bound is saturated if and only if $F^+_H=0$ or $F^{-}_H=0$, together with $F_V=0$. We will refer to these conditions as a contact instanton. 
  The bound (\ref{full-5D-bound}) can be written for any contact manifold $M_5$ and it represents an analog of the
   standard $4D$-bound for the Yang-Mills action. However there is an important difference, since the term on the right hand side of 
    (\ref{full-5D-bound}) is not a proper topological term. 

 The conditions for a contact instanton  $F^\pm_H=0$ and $F_V=0$ can be equivalently rewritten as follows
  \begin{equation}
   \iota_v (* F) = \mp  F~.\label{cov-5D-inst}
  \end{equation}
  Alternatively, equation (\ref{cov-5D-inst})  can also be written as
    \begin{equation}
     \kappa \wedge F = \mp *F~.
  \end{equation}
   The above equation  implies $\iota_v F=0$ and it  corresponds to a ``normal" instanton in the contact plane. 
   In the next section we discuss the 5D theory which localizes on contact instantons.  
   
   We have  little intuition about contact instantons, but it seems that they are not a trivial generalization of
    4D instantons. As an example, let us consider the canonical contact structure $\kappa$ on $\mathbb{R}^5$
    \begin{equation}
    \kappa = dz + x_1 dy_1 + x_2 dy_2~,\label{can-contstr-5D}
    \end{equation}
    where $(x_1, y_1, x_2, y_2, z)$ are the coordinates on $\mathbb{R}^5$. The Reeb vector field is $v = \partial_z$ and 
     $\kappa \wedge d\kappa \wedge d\kappa$ is a canonical constant volume form on  $\mathbb{R}^5$. However, 
      the metric compatible with the contact structure is not a standard flat metric, but the following metric
      \begin{equation}
       g= dx_1 \otimes dx_1 + dy_1\otimes dy_1 + dx_2 \otimes dx_2 + dy_2 \otimes dy_2 +  \kappa \otimes \kappa~.\label{metric-con5D}
      \end{equation}
     This metric is non-constant and the corresponding contact instanton equations do not reduce to a 4D problem.
      In any case, these equations will require further proper study and we leave it for the future.  
      
\subsection{5D Yang-Mills theory and localization}
\label{5DYM-sub}
The 5D version of the transformations (\ref{general-transform}) we will use in order to localize our theory to contact instantons are given by
  \begin{equation}\label{5D-transform}
  \begin{split}
 &\delta A = \Psi~, \\
 & \delta \Psi = \iota_v F + i d_A \sigma~, \\
 & \delta \sigma = - i~ \iota_v \Psi~,\\
 & \delta \chi_H^+ = H_H^+~,  \\
 & \delta H_H^+ =  {\cal L}^A_v \chi_H^+- i [\sigma, \chi_H^+]~,
 \end{split}
 \end{equation}
where $\chi_H^+$ is a fermonic two-form in $\Omega_H^{2+}(M_5)$ and $H^+_H$ is a bosonic two-form, also in $\Omega_H^{2+}(M_5)$. There is one tricky point here. Unless we assume that ${\cal L}_v$ commutes with the Hodge-star (namely, ${\cal L}_v g=0$ so that $v$ is an isometry for the metric $g$), ${\cal L}^A_v \chi_H^+$ may not be in $\Omega_H^{2+}(M_5)$. However, assuming that ${\cal L}_v g=0$, we have consistent transformations (\ref{5D-transform}) which square to ${\cal L}_v$ plus a gauge 
  transformation.
  
On a contact manifold $M_5$ with the choice of a compatible metric we can write the following BRST-exact term
\begin{equation}
S_{SYM_5} = \int\limits_{M_5} {\rm Tr} \left (4 \delta (\chi_H^+ \wedge *( F^+_H - \frac{1}{2} H_H^+) + \Psi \wedge * \overline{\delta \Psi}) \right )~.
\label{BRST-5D-exact}
\end{equation}
 Next we would like that the BRST-exact action (\ref{BRST-5D-exact}) is BSRT-invariant. For this again we need
   the condition that  ${\cal L}_v g=0$ and thus the Lie derivative ${\cal L}_v$ goes through the Hodge star and we get $\delta S_{SYM_5}=0$.
Working out explicitly the action (\ref{BRST-5D-exact}) and integrating out $H_H^+$ we arrive at the following action       
\begin{equation}
\begin{split}
S_{SYM_5} = \int\limits_{M_5} &{\rm Tr}  (F \wedge * F + \kappa \wedge F \wedge F + d_A \sigma \wedge * d_A\sigma + 2\chi_H^+\wedge *\left(\mathcal{L}^{A}_v\chi_H^+-[i\sigma,\chi_H^+]\right) \\
 &-\Psi\wedge * \left(\iota_v d_A \Psi-d_A\iota_v \Psi +[i\sigma,\Psi]\right)-4\chi_H^+\wedge * d_A \Psi)~,\label{5D-final-YM}
\end{split}
\end{equation}
where we have used the relation (\ref{5D-YM-H-V}). Thus, the action (\ref{5D-final-YM}) is a twisted version of supersymmetric 5D Yang-Mills theory. The field content of this theory can be identified with the field content of physical $N=1$ 5D supersymmetric Yang-Mills theory. The fermionic fields $(\Psi, \chi_H^+)$ in the present model have $5+3=8$ components which is the same 
   as a Dirac spinor in 5D.  We will comment  more on the relation between physical $N=1$ SYM on  ${\mathbb R}^5$ and the present twisted supersymmetric Yang-Mills in subsection \ref{flat-5DYM}. 
  
   By standard localization arguments, the action $\eqref{BRST-5D-exact}$ can be used to localize the 5D theory to solutions to the equations
   \begin{equation} \label{5dmod}
   \begin{split}
   \iota_v*F&=- F~, \\
   d_A\sigma &= 0~.
   \end{split}
   \end{equation}

  \subsection{Chern-Simons type observables} 
\subsubsection{Lifting 3D term to 5D}
\label{3D-5D}

Since on a contact manifold $M_5$ there is a two form $d\kappa$, we can lift the 3D Chern-Simons action (\ref{classical-CS}) to 5D in the 
 following trivial fashion 
 \be
 S_{CS_{3,2}} (A)=  \int\limits_{M_5} {\rm Tr} \left ( d \kappa \wedge  ( A \wedge d A + \frac{2}{3} A \wedge A \wedge A  ) \right )~,
 \ee  
which can be equivalently written as
  \be
 S_{CS_{3,2}} (A)= \int\limits_{M_5} {\rm Tr} \left ( \kappa \wedge  F \wedge F \right ) ~.
\ee  
 Now we can put any coupling in front and it does not require any quantization condition. 
The supersymmetric version of $S_{CS_{3,2}}$ is a simple lift of the 3D action (\ref{shift-3D-action}) to 5D,
\be
 S_{SCS_{3,2}} = S_{CS_{3,2}} (A-i \sigma \kappa) - \int\limits_{M_5} {\rm Tr} \left (d\kappa \wedge \kappa \wedge \Psi \wedge \Psi \right )~, 
\ee
 which can be rewritten as follows    
 \begin{equation} \label{cs3obs}
  S_{SCS_{3,2}} = \int\limits_{M_5} {\rm Tr} \left  ( \kappa \wedge F \wedge F - d\kappa \wedge \kappa \wedge \Psi \wedge \Psi - 
   2i d\kappa \wedge \kappa \wedge F \sigma - \sigma^2 \kappa \wedge (d\kappa)^2 \right )~.
 \end{equation}
 One can easily check that this action is invariant under the transformations (\ref{5D-transform}), and the properties of the contact structure plays
  a crucial role for this invariance. In addition, the above action has the shift symmetry $\eqref{shift}$.
\subsubsection{5D Chern-Simons}
\label{5D-CS-sub}

Next consider the standard five-dimensional Chern-Simons action
\be
 S_{CS_5} (A) =\frac{\rm k}{24\pi^2} \int\limits_{M_5} {\rm Tr } \left (A\wedge dA \wedge dA + \frac{3}{2} A\wedge A \wedge A \wedge dA + \frac{3}{5} A\wedge 
  A \wedge A \wedge A \wedge A \right )~,
\ee
 where ${\rm k} \in {\mathbb Z}$ according to the standard arguments. Following our previous discussion it is easy to 
  write the supersymmetric extension of this action:
 \begin{equation} \label{cs5obs}
 S_{SCS_5} = S_{CS_5} (A - i\sigma \kappa) - \frac{\rm k}{ 8 \pi^2} \int\limits_{M_5} {\rm Tr} \left ( \Psi \wedge \Psi \wedge \kappa \wedge F(A-i\sigma \kappa)
  \right )~.
\end{equation}
  This action is invariant under the transformations $\eqref{5D-transform}$ and it has a shift symmetry.  We can construct similar 
   supersymmetric Chern-Simons actions in higher dimensions, see Appendix \ref{GCS-terms} for explicit expressions.
  
 There is a crucial difference between the two observables $\eqref{cs3obs}$ and $\eqref{cs5obs}$. The latter one depends only on the contact structure, but not on the concrete contact form $\kappa$. This is since the action (\ref{cs5obs})  
  has the following symmetry
  \be
  A~\rightarrow~A~,~~~\kappa~\rightarrow~ e^f \kappa~,~~~\Psi~\rightarrow~e^{-f/2} \Psi~,
  ~~~\sigma~\rightarrow~e^{-f}\sigma~, 
  \ee
and thus it depends only on 
  the contact planes, ${\rm ker}(\kappa) \subset T M_5$. This is not true for the action $\eqref{cs3obs}$, and consequently this observable is not topological since it depends on the chosen contact form.  
  
  Now by having different supersymmetric actions $S_{SYM_{3,2}}$, $S_{SCS_5}$ and $S_{SYM_5}$ we can combine them and 
  get other invariant actions. For example, we can study the combination 
\be
 S_{SCS_{3,2}} - S_{SYM_5} = - \int\limits_{M_5} {\rm Tr} ( F \wedge * F +  d_A \sigma \wedge * d_A\sigma + 8 \sigma^2 {\rm vol}_g
  + 2i d\kappa \wedge \kappa \wedge F \sigma + \cdots )~, 
\ee
 where the dots stand for the terms involving fermionic fields. One can proceed further and construct other terms and actions which 
  will be invariant under twisted supersymmetry.

\subsection{Aspects of contact geometry}

In the previous subsections when we were constructing the theory we came up with a number of conditions on the contact manifold $M_5$.
 Namely, we need a contact manifold $M_5$ with a metric $g$ such that this metric is compatible with the contact structure 
 ($\iota_v ( * \omega_p) = (-1)^p * (\kappa \wedge \omega_p)$) and moreover we have to require that ${\cal L}_v g=0$. 
  The last condition indicates that the Reeb vector field $v$ should correspond to a $U(1)$-action. Thus the natural question to ask 
   is if there are any examples of five manifolds $M_5$ where all these conditions are satisfied. Indeed there are a plenty examples, given 
    by $U(1)$-bundles over symplectic four manifolds with an integral symplectic form.

Let us explain this construction which is due to a well-known theorem of Boothby and Wang \cite{MR0112160}.  Consider a $S^1$ bundle 
\be
\begin{array}{lll}
  M_{2n+1} & \longleftarrow & S^1\\
{\scriptstyle \pi}\Big\downarrow && \\
  ~\Sigma_{2n} &   & 
\end{array}\label{bundle-general-BW}
\ee
 over a symplectic manifold $\Sigma_{2n}$ with a symplectic form $\omega$ such that  $\omega \in H^2(\Sigma_{2n}, \mathbb{Z})$. 
  The Boothby-Wang theorem states that we can choose the contact structure $\kappa \neq 0$ on $M_{2n+1}$ such that 
  $\kappa$ is a connection one-form on this bundle $M_{2n+1}$ and $d\kappa = \pi^* (\omega)$.  The Reeb vector field $v$
   corresponds to the $S^1$-action on $M_{2n+1}$.  Therefore it is easy to construct an invariant metric with the required property. 

A simple example of this theorem is given by the Hopf fibration
\be
\begin{array}{lll}
  S^{2n+1} & \longleftarrow & S^1\\
{\scriptstyle \pi}\Big\downarrow && \\
  {\mathbb C}P_{2n} &   & 
\end{array}\label{bundle-general-BW}
\ee
 On ${\mathbb C}P_{2n}$ we choose the standard symplectic form associated to the Fubini-Study K\"ahler metric.
  The induced contact structure on $S^{2n+1}$ is called the standard contact structure on the unit sphere $S^{2n+1}$. 
   If we think of $S^{2n+1}$ as a unit sphere in ${\mathbb R}^{2n+2}$, then upon choosing Cartesian coordinates 
  $(x_1, y_1, ~\cdots , x_{n+1}, y_{n+1})$ on ${\mathbb R}^{2n+2}$ we define the one-form
  \be
   \tau= \sum\limits_{j=1}^{n+1} ( x_j dy_j - y_j dx_j)~.
  \ee
   This one-form induces the standard contact form $\kappa$ on unit sphere $S^{2n+1}$. This contact form is a connection 
    for the Hopf fibration.  The orbits of the corresponding Reeb vector field are the fibers of this fibration.  
 In section \ref{calculation}  we will do an explicit calculation for the case of $S^5$, which is understood as $S^1$-fibration over ${\mathbb C}P_2$.     

 One can construct many more examples, by choosing $\Sigma_{2n}$ to be a K\"ahler manifold with integral K\"ahler form, so that 
  $\Sigma_{2n}$ is a Hodge manifold that can be realized as an algebraic variety in projective space.

 \subsection{On twisted supersymmetric Yang-Mills in the flat case}
 \label{flat-5DYM}
 
 In this subsection we would like to comment on the relation with the standard $N=1$ 5D supersymmetric Yang-Mills and related 
  issues. 
  
  If we consider the case when $M_5 = \Sigma_4 \times S^1$ is a trivial $S^1$ bundle, then beside using a contact structure we can 
  use something weaker. Namely, introducing a coordinate $t$ along $S^1$ we can take $\kappa=dt$ and $v = \partial_t$.  In this 
   case $d\kappa =0$ identically. All of our previous discussions goes through for this situation, since we only need to use the properties $\iota_v \kappa=1$
    and $\iota_v d\kappa=0$ in subsections \ref{3D-5D} and \ref{5D-CS-sub}. However, now the terms with $d\kappa$ in the different expressions are identically 
     zero. In subsections \ref{gauge-fix-sub} and \ref{5DYM-sub} we should use now the obvious notion of vertical 
    and horizontal decomposition given by $\kappa=dt$ and $v=\partial_t$. Indeed, this situation was discussed in Nekrasov's work \cite{Nekrasov:1996cz} and
     in the Baulieu-Losev-Nekrasov work \cite{Baulieu:1997nj}. With the choice $\kappa=dt$ and $v=\partial_t$ the compatible metric $g$ is the block diagonal one and the condition ${\cal L}_v g=0$ says that it is independent of the $S^1$-coordinate $t$. For the case $M_5 = {\mathbb R}^4 \times S^1$ (or $M_5 = {\mathbb R} \times {\mathbb R^4}$) with the flat metric we will recover the theory which can be thought as a topological twist of the $N=1$ supersymmetric 
         Yang-Mills action. The action will be the same and the field contend can be easily mapped, since a Dirac fermion can be decomposed
          into $\Psi$ and $\chi_H^+$.  

Even in the case of ${\mathbb R}^5$ we can consider the case of canonical contact structure (\ref{can-contstr-5D})
 and take a compatible non-flat metric metric (\ref{metric-con5D}). In this case the BRST-exact term (\ref{5D-final-YM}) will correspond to 
  5D Yang-Mills theory with a non-flat metric and the supersymmetric extensions of the Chern-Simons actions are different compared 
   to the case discussed above. Moreover, unlike the $3D$ case, the contact structure enters more drastically in the $5D$ story. For example, the localization equations $\eqref{5dmod}$ depends on the contact structure. In the general 5D case, it is an interesting question how much the partition function and expectation values of different observables feel the choice of contact structure. We leave this subject for future studies.  

The main point of the above discussion is that on manifolds which has a product structure, for example $M_5=S^{1}\times M_4$, we can have different choices of $\kappa$ and $v$ which fulfills our requirements to have a consistent theory. Choosing $\kappa=d t$ and $v=\frac{\partial}{\partial t}$, our construction is the same as in \cite{Nekrasov:1996cz}, whereas choosing $\kappa$ and $v$ to be a contact structure and the Reeb vector field we enter unexplored territory. Moreover, the construction with a contact structure has an obvious generalization for non-trivial $S^1$-bundles.   
\section{Partition function on $S^5$}
\label{calculation}

Our goal is to calculate the full perturbative partition function for the theory
\begin{equation}\label{Z5dtheory}
Z(w, {\rm k}, s)= \int {\rm exp} \left (  i w S_{SCS_{3,2}} + iS_{SCS_5} + s S_{SYM_5} \right) 
\end{equation}
 defined over $S^5$. The final result will be presented in terms of a matrix model. Due to standard arguments the partition 
  function will be independent of $s$, since it enters in front of a $\delta$-exact term. We calculate $Z$ by sending $s$ to infinity together with standard localization arguments. 
   Here $s$ carries dimension, so we should use the radius $R$ of the sphere and write $s= \tilde{s} R^{-1}$. Thus we will send 
    to infinity the dimensionless parameter $\tilde{s}$. 
   The parameter $w$ here is also dimensionfull and it will introduce a dependence of the radius of $S^5$. To avoid cluttering 
    in the formulas we set $R=1$ and we can easily reintroduce the $R$-dependence in the final result. 
     If $w=0$ then we calculate the partition function of a supersymmetric version of $5D$ Chern-Simons theory, which is supposed to be a topological field theory.

\subsection{Reminder of localization}
In general, the calculation of the expectation values of $\delta$-closed, gauge invariant observables can be computed in the standard way by using the localization technique for path integrals. This technique is an infinite dimensional generalization  \cite{Witten:1988ze,Witten:1991zz} of the Atiyah-Bott-Berline-Vergne localization theorem \cite{Atiyah:1984,Berline:1982}, which we for completeness now briefly review. Let $M$ be a manifold which admits an action of a compact group $\mathcal{H}$, and let the action on $M$ be generated by the vector field $v$. We then consider equivariant differential forms, that is, differential forms with values in the Lie algebra of $\mathcal{H}$, invariant under the action of $v$. We form the equivariant differential $Q=d-\phi^{a}i_{v^{a}}$, where $d$ is the de Rham differential, $\phi^{a}$ is a parameter for the action. We have $Q^{2}=-\phi^{a}\mathcal{L} _{v^{a}}$, and hence $Q$ is a differential when acting on equivariant differential forms. Consider now a $Q$-closed equivariant form $\alpha$. The localization formula says that 
\be
\int\limits_{M}{\alpha}=\int\limits_{Y}{\frac{i^{*}_{Y}\alpha}{e(N_Y)}}~,
\ee
where $Y\subset M$ is the set of zeros of the vector field $v$, and $e(N_Y)$ is the equivariant Euler class of the normal bundle of $Y$. In the case where $Y$ is a set of discrete points, the formula reduces to 
\begin{equation}\label{abbvdis}
\int\limits_{M}{\alpha}=\sum_{p\in Y}{{\frac{i^{*}_{p}\alpha}{\sqrt{\text{det}~ L_p}}}}~,
\end{equation}
where $L_p$ is the action of $\mathcal{H}$ on the tangent space at the point $p$.

In our situation, $M$ corresponds to the space of fields and $\alpha$ is any product of the observables defined in section \ref{5D gauge theory}. As we will see below, after gauge fixing, the group $\mathcal{H}$ will correspond to $U(1)\times G$, where $G$ denotes the group of constant gauge transformations. The subset $Y$ of $M$ is by construction given by the solution to the set of equations
\begin{equation} \label{fpset}
\begin{split}
\iota_v* F &= - F~, \\
d_A\sigma & = 0~,
\end{split}
\end{equation}
modulo gauge transformations.  

On $S^5$, the trivial connection is an isolated point in the space of solutions to the equations $\eqref{fpset}$, since the first cohomology group of $S^5$ is trivial. We will below compute the contribution to the expectation value of the observables defined in section \ref{5D gauge theory} coming from this point in the space of solutions to $\eqref{fpset}$. The computation we will do is very similar to the ones performed in \cite{Kallen:2011ny,Beasley:2005vf,Blau:2006gh,Pestun:2007rz}, with a matrix model as the end result. However, in the 5D theory the one-loop determinant will give an extra factor in the matrix model as compared to the 3D theory. We will be rather brief below and only explain the main points and giving the end result of the calculation. For details, we refer to appendix \ref{1loopcalc}.
\subsection{Gauge fixing}
Before we can do calculations, we must gauge fix the theory. As shown in \cite{Baulieu:1997nj,Pestun:2007rz}, we can always introduce the standard set of ghost fields together with BRST transformations, and combine this BRST symmetry with the transformations $\delta$ defined in
  $\eqref{5D-transform}$. This combined transformation is denoted $Q$ below.  We introduce the set of ghost fields $c,\bar{c},b,a_0,\bar{a}_0,\bar{c}_0,b_0,c_0$ and transformations
\begin{align}\label{Q}
QA&=\Psi+d_Ac~, \nonumber \\
Q\Psi&= i_v F + i d_A \sigma +[\Psi,c]~, \nonumber\\
Q\chi_H^+& = H^+_H+[\chi_H^+,c] ~,\nonumber\\
QH_H^+ &=  {\cal L}^A_v \chi_H^+ - i [\sigma, \chi_H^+] + [H_H^+,c]~,  \\
Qc&=a_0- (i \sigma - i_v A)-\frac{1}{2}[c,c]~, & Q\bar{c}&=b~, & Q\bar{a}_0&=\bar{c}_{0}~, & Qb_0&=c_0~,\nonumber \\
Q\sigma&=-i~i_v\Psi+G_c\sigma~, & Qb&=\mathcal{L}_v\bar{c}+G_{a_0}\bar{c}~, & Q\bar{c}_0&=G_{a_0}\bar{a}_0~, & Qc_0&=G_{a_0}b_0~.\nonumber  \\
Qa_0&=0 \nonumber
\end{align}
The fields  $c,\bar{c},b$ are the standard ghost and Lagrange multiplier fields, whereas $a_0,\bar{a}_0,\,b_0$ are even zero modes and $c_0,\bar{c}_0$ are odd zero modes. $Q$ fulfills $Q^{2}=\mathcal{L}_v+G_{a_0}$ on all fields. As shown in \cite{Kallen:2011ny,Pestun:2007rz}, we can impose the gauge condition
\begin{equation}
d^{\dagger}A=0~,
\end{equation}
where $d^{\dagger}$ is the adjoint operator to $d$ using the inner product defined by the Hodge star, by adding a $Q$-exact term to the action in the standard way.   

\subsection{Summary of calculation on $S^5$}
We can interprete $Q$ as an equivariant differential acting on a supermanifold with even coordinates $(A, \bar{a}_0,b_0)$ and odd coordinates $(\chi_H^+, c, \bar{c})$. The group used to define the equivariant differential is $U(1)\times G$. The parameter for the $G$ transformations is a field in our theory, denoted by $a_0$, which we integrate over separately. We want to compute the gauge fixed version of $\eqref{Z5dtheory}$. The gauged fixed version means that there is a gauge fixing term added to the integrand and the integral is over all fields appearing in $\eqref{Q}$. Using localization, this infinite dimensional integral reduce to an integral over the space of solutions to the equations $\eqref{fpset}$. On $S^5$, the trivial connection is an isolated point in this space and we can split up $Z$ into contributions coming from this point and the rest:
\be
Z=Z_{\{0\}}+Z_{np}~.
\ee
Below we will compute the part $Z_{\{0\}}$. 

For the trivial connection, the solution to the second equation in $\eqref{fpset}$ is given by
\be
\sigma = {\rm constant}~,
\ee
that is, $\sigma$ is a constant field taking value in the Lie algebra $\mathfrak{g}$. Putting together all ingredients, it was shown in detail in \cite{Kallen:2011ny} that $Z_{\{0\}}$ will be given by the integral
\begin{equation} \label{Z0}
Z_{\{0\}}=\frac{1}{|W|}\frac{\text{Vol(G)}}{\text{Vol(T)}}\int\limits_{\mathfrak{t}}{[d\phi]~\prod_{\beta>0}{\langle\beta,\phi\rangle^2}~\alpha(\phi)~h(\phi)}~,
\end{equation}
where the integral is over the Cartan subalgebra $\mathfrak{t}$ of the Lie algebra $\mathfrak{g}$. $\beta$ denotes the roots of the Lie algebra $\mathfrak{g}$, and $\langle~,~\rangle$ denotes the natural pairing between $\mathfrak{g}$ and $\mathfrak{g}^*$. $\alpha(\phi)$ denotes the integrand in $\eqref{Z5dtheory}$ evaluated at 
\begin{equation}
\begin{split}
&\sigma={\rm constant}~,\\
&\text{All other fields}=0~,
\end{split}
\end{equation}
and we have denoted $i\sigma=\phi$. $|W|$ denotes the order of the Weyl group of $G$, $T$ denotes the maximal tours of $G$, $h(\phi)$ denotes the one-loop determinant. In appendix \ref{1loopcalc} we show that $h(\phi)$ is given by
\be
&&h(\phi)=e^{\frac{i\pi}{8}\Delta_G+\check{c}_{\mathfrak{g}}\frac{3i}{8\pi}\text{Tr}\phi^{2}}\cdot e^{-\Delta_G \zeta'(-2)}\cdot \prod_{\beta>0}{\left[\frac{\sin(\pi x)}{\pi x}\right]^2\cdot e^{f(x)}}~.
\ee
Here $\check{c}_{\mathfrak{g}}$ is the dual Coxeter number of $\mathfrak{g}$, $\zeta$ is the Riemann zeta-function and
\begin{equation}\label{xfx}
\begin{split}
x& =\frac{\langle \beta,\phi\rangle}{2\pi}~,\\
f(x)&=\frac{i\pi }{3}x^3+x^2\text{ln}(1-e^{-2\pi i x})+\frac{i}{\pi}x\text{Li}_2(e^{-2\pi i x})+\frac{1}{2\pi^2}\text{Li}_3(e^{-2\pi i x}) -\frac{\zeta(3)}{2\pi^2}~ .
\end{split}
\end{equation}
In order to calculate $\alpha(\phi)$, we need to evaluate the observables defined in $\eqref{cs3obs}$ and $\eqref{cs5obs}$ at the trivial solution to $\eqref{fpset}$. We find
\begin{equation}
\begin{split}
w~\text{Tr}\int{\kappa \wedge d\kappa \wedge d\kappa ~\phi^2}&=w~\text{Tr}\phi^2~, \\
\frac{k}{24\pi^2}\text{Tr}\int{\kappa \wedge d\kappa \wedge d\kappa~ \phi^3}&=\frac{k}{24\pi^2} \text{Tr}\phi^3~,
\end{split}
\end{equation}
and the integrand in $\eqref{Z5dtheory}$ is given by
\begin{equation}
\alpha(\phi)=\exp\left[i w ~\text{Tr}\phi^{2}+\frac{ik}{24\pi^2}\text{Tr}\phi^3\right].
\end{equation}
The final expression for $\eqref{Z0}$ is thus given by
\begin{equation}
Z_{\{0\}}= A \int_{\mathfrak{t}}{[d\phi]~\exp\left[\left(iw+\check{c}_{\mathfrak{g}}\frac{3i}{8\pi}\right)\text{Tr}\phi^{2}+\frac{ik}{24\pi^2}\text{Tr}\phi^3\right]\cdot \prod_{\beta>0}{\left[2\sin(\pi x)\right]^2\cdot e^{f(x)}}}~,\label{final-matrix-model}
\end{equation}
where $A$ is a constant 
\be
   A= \frac{1}{|W|}\frac{\text{Vol(G)}}{\text{Vol(T)}}\cdot e^{\frac{i\pi}{8}\Delta_G-\Delta_G \zeta'(-2)}
\ee
 and $x$ and $f(x)$ are defined in $\eqref{xfx}$.

 Let us make a few final remarks on the answer (\ref{final-matrix-model}). It is possible to introduce all parameters in the 
  final answer. Assuming that we have a sphere $S^5$ of radius $R$ and $S_{SCY_{3,2}}$ has a dimensionfull parameter $w$ then 
   the final answer, modulo overall factors, is  
   \be
   Z_{\{0\}} \sim \int_{\mathfrak{t}}{[d\phi]~\exp\left[i \left(wR+\check{c}_{\mathfrak{g}}\frac{3}{8\pi}\right)\text{Tr}\phi^{2}+\frac{ik}{24\pi^2}\text{Tr}\phi^3\right]\cdot \prod_{\beta>0}{\left[2\sin(\pi x)\right]^2\cdot e^{f(x)}}}~,
\ee
where $\phi$ now is a dimensionless combination. We remember that the field $\sigma$ carries dimension and we should use 
 $R$ to get a dimensionless combination.  If we set $w=0$, then we get the perturbative partition function for the supersymmetric version of 5D Chern-Simons theory
  and there is no dependence on $R$, as expected since this is a topological field theory. 
  
At the present moment, we cannot say much about properties of this matrix model. We expect that it can be helpful to understand properties of perturbative five-dimensional supersymmetric Yang-Mills theory, but this aspect of the present work requires further study. We plan to come back to this elsewhere.

\section{Summary and outlook}
\label{end}

In this work we have constructed and studied 5D gauge theories which are invariant under the twisted supersymmetry 
 transformations (\ref{5D-transform}). Contact geometry plays a crucial role in our construction. In particular, we have constructed
  a twisted 5D Yang-Mills theory with different Chern-Simons couplings defined over $S^1$-fibrations of four dimensional symplectic 
   manifolds with integral symplectic form. The field content of our theory can be mapped to the field content of a five-dimensional
    supersymmetric gauge theory with eight supercharges, which in five dimensions is $N=1$ supersymmetric Yang-Mills theory. The study of $N=1$ 
    supersymmetric Yang-Mills theory was initiated in \cite{Seiberg:1996bd} and have attracted some attention during the last years. 
   Our theory can be understood as some version of a topological twist for this physical theory. The implications of our study for the physical 
    theory are unclear to us at the moment.    

  Our work is a generalization and covariantization of previous works by Nekrasov \cite{Nekrasov:1996cz}, by Baulieu, Losev and Nekrasov \cite{Baulieu:1997nj} and of earlier works
   by  Nair and Schiff  \cite{Nair:1990aa, Nair:1991ab}.  However, 
   there is an important difference in that our construction can handle non-trivial $S^1$-bundles over symplectic four manifolds. 
    An example of such a manifold is $S^{5}$, and for this manifold we have calculated the full perturbative partition function. It is difficult to produce the full 
     non-perturbative answer since this requires  an understanding of how to handle instantons over ${\mathbb C}P^2$. 
   Our perturbative calculation can be 
     generalized straightforwardly to other $S^1$-fibrations with trivial $H^1(M_5, {\mathbb R})$, so that the trivial connection is an isolated flat connection. With non-trivial $H^1(M_5, {\mathbb R})$ our calculation should be modified. 

  Beside concentrating on $S_{SYM_5}$, our calculation is also related to two Chern-Simons theories in 5D, 
  $S_{CS_{3,2}}$ and $S_{CS_5}$. For trivial bundles these theories were discussed in the above mentioned papers and 
   in particular there is an interesting analog of  four dimensional ${WZW}$-theory discussed in \cite{Losev:1995cr}.  Unfortunately, not much 
    is known about 5D Chern-Simons theory. Even the perturbation theory is not straightforward to construct and study. 

    There are many interesting results in contact geometry which should have interesting impact on generalizations and extensions 
     of our analysis. Let us mention one example of this situation. 
 In the future we indent to study the present model on $M_3 \times T^2$. This manifold admits a $T^2$-invariant contact structure \cite{MR1912277}.
  For the physical $N=1$ Yang-Mills theory, a similar analysis was performed in \cite{Haghighat:2011xx} for the case when 
   $M_3 = {\mathbb R}^3$.  In particular, it would be interesting to study $S_{CS_{3,2}}$ on $M_3 \times T^2$. There is no 
    quantization condition for  $S_{CS_{3,2}}$ and moreover, the final answer for this 5D model should behave nicely under the modular group
     of $T^2$. This may lead to interesting invariants of $M_3$. 

\bigskip\bigskip

\noindent{\bf\Large Acknowledgement}:
\bigskip

\noindent  We thank Tobias Ekholm, Alexander Gorsky, Sergei Gukov,  Vasily Pestun and Jian Qiu  for useful discussions 
 on this and related subjects.  
 We are grateful to Patrick Massot for pointing out a mistake in an earlier version of the draft.  We thank the referee for valuable comments 
  on   the paper. 
  The research of MZ is supported by VR-grant 621-2011-5079.
\appendix
 
\section{Summary of contact geometry}
\label{geometry-cont}

An odd dimensional manifold $M_{2n+1}$ is called a contact manifold if it admits  a contact
structure which is a maximally non-integrable hyperplane field $\xi ={\rm ker} ~\kappa \subset T M_{2n+1}$. The differential one-form $\kappa$ is required to satisfy
$$ \kappa \wedge (d\kappa)^n \neq 0~.$$
 Such a one-form $\kappa$ is called a contact
form. The pair $(M_{2n+1}, \xi)$ is called a contact manifold. We can always define a vector field $v$
 with properties $\iota_v \kappa=1$ and $\iota_v d\kappa=0$.  Such a vector field $v$ is called the Reeb vector 
  field. We can change $\kappa \rightarrow e^f \kappa$,   where $f$ is a function on 
   $M_{2n+1}$. Under such a change the contact plane $\xi ={\rm ker }~\kappa$ does not change  and this is understood as
    the same contact structure.  However, the Reeb vector field will change in rather complicated fashion. 
    Contact geometry and contact topology is a booming area of research, see \cite{MR2397738}
     for a nice modern introduction to the subject. 

  On a contact manifold $M_{2n+1}$ we can always choose a metric $g$ compatible with the contact structure such that 
   the following is satisfied 
$$\iota_v ( * \omega_p) = (-1)^p * (\kappa \wedge \omega_p)~,$$
where $\omega_p$ is $p$-form and $*$ is the Hodge star operation corresponding to the metric $g$. The above property 
 can be easily written in local coordinates as follows $\kappa_{\mu} = g_{\mu\nu} v^\nu$.  The volume forms corresponding 
  to the metric $g$ and the contact structure are related as follows
$${\rm vol}_g = \frac{(-1)^n}{2^n n!} \kappa \wedge (d\kappa)^n~.$$
 For the proof of the existence of a compatible metric and other related properties the reader may consult the textbook
\cite{Blair:2010}. 

\section{General Chern-Simons terms}
\label{GCS-terms}

Let us discuss the case of supersymmetric Chern-Simons terms in higher dimensions. 
For concreteness we discuss
 only the 7D and 9D cases. Further generalization to higher dimensions are straightforward. 
The standard Chern-Simons action on  $M_{2n+1}$ can be defined as follows
\be
S_{CS_{2n+1}}(A) =(n+1)\int\limits_{M_{2n+1}}  \left [ \int\limits_{0}^{1}{ds~s^n {\rm{Tr}}\left(A\wedge(d A+s A\wedge A)^n\right)} \right ]~,
\ee
 where we introduce the auxiliary parameter $s$. 
 On a seven dimensional manifold $M_7$ we can construct the following observable
\be
&&S_{SCS_7} = S_{CS_{7}}(A-i\sigma \kappa)-  \int\limits_{M_7}{\rm{Tr}}\Big( 4[ \Psi\wedge\Psi\wedge F(A-i\sigma \kappa) \wedge F(A-i\sigma \kappa)    \nonumber \\
&&   + \frac{1}{2}\Psi\wedge F(A-i\sigma \kappa)\wedge \Psi\wedge F(A-i\sigma \kappa) ] \wedge \kappa- ~\Psi\wedge \Psi\wedge \Psi\wedge \Psi\wedge \kappa\wedge d\kappa \Big )
\ee
and on a nine dimensional manifold $M_9$ have the observable
\be
&&S_{SCS_9}=  S_{CS_{9}}(A-i\sigma \kappa)-   \int\limits_{M_9}{\rm{Tr}}\Big( 5 [\Psi\wedge\Psi\wedge F(A-i\sigma \kappa) \wedge F(A-i\sigma \kappa)\wedge F(A-i\sigma \kappa)\nonumber \\
&&+\Psi\wedge F(A-i\sigma \kappa)\wedge F(A-i\sigma \kappa)\wedge \Psi\wedge F(A-i\sigma \kappa)] \wedge\kappa\nonumber \\
&&-5~\Psi\wedge \Psi\wedge \Psi\wedge \Psi\wedge F(A-i\sigma \kappa) \wedge \kappa\wedge d\kappa\Big)~.
\ee
 The actions $S_{SCS_7}$ and $S_{SCS_9}$ are invariant under the transformations (\ref{general-transform}) assuming that $\kappa$ is a contact form
  and $v$ is the corresponding Reeb vector field.  Moreover the actions $S_{SCS_7}$ and $S_{SCS_9}$   have an additional gauge shift-symmetry
     \be
     \tilde{\delta} A = \xi \kappa~,~~~~~~~~\tilde{\delta} \sigma = - i \xi~.
     \ee

\section{Computation of the 1-loop determinant}
\label{1loopcalc}
The computation of the 1-loop determinant goes along the same lines as the computations in \cite{Kallen:2011ny,Beasley:2005vf,Blau:2006gh,Pestun:2007rz}. As explained in the main text, the transformations $\eqref{Q}$ can be interpreted as an equivariant differential acting on a supermanifold with even coordinates $(A,\bar{a}_0,b_0)$ and odd coordinates $(\chi,c,\bar{c})$. In this appendix we will compute the determinant appearing in $\eqref{Z0}$, namely the determinant of the operator 
\begin{equation}
L_{\phi}:=\mathcal{L}_v+G_{\phi}
\end{equation}
acting on the tangent space of the space of fields. $\mathcal{L}_v$ is the Lie derivative in the direction of the Reeb vector field, and $G_\phi$ denotes the adjoint action with parameter $\phi\in\mathfrak{t}$. This determinant will be a function of $\phi$, and it is denoted by $h(\phi)$.

The tangent space to the bosonic part of the space of fields is given by 
\begin{equation}
\Omega^{1}(S^{5},\mathfrak{g})\oplus H^{0}(S^5,\mathfrak{g})\oplus H^{0}(S^5,\mathfrak{g})~,
\end{equation}
where $H^{0}(S^5,\mathfrak{g})$ denotes the space of harmonic zero forms with values in the Lie algebra $\mathfrak{g}$. The fermonic part of the tangent space of the space of fields is given by
\begin{equation}
\Omega^{2+}_H(S^5,\mathfrak{g})\oplus\Omega^{0}(S^5,\mathfrak{g})\oplus\Omega^{0}(S^5,\mathfrak{g})~,
\end{equation}
where $\Omega^{2+}_H$ denotes the space of horizontal, self-dual two-forms. Since $S^5$ is a $S^1$ fibration over $\mathbb{C}P^2$ with connection $\kappa$, and $\mathbb{C}P^2$ is a K\"ahler manifold, we can use these geometric structures to decompose the space of differential forms. Choosing a complex structure and a symplectic structure, we can  we can decompose the spaces $\Omega^{1}$ and $\Omega^{2+}_{H}$ as 
\begin{equation}
\begin{split}
\Omega^{1}=&\Omega_V \oplus\Omega^{1}_{H}~, \\
\Omega_{H}^{2+}=&\Omega^2_{\omega}\oplus\Omega_{H}^{(2,0)}\oplus\Omega^{(0,2)}_H~, 
\end{split}
\end{equation}
where $\Omega_V$ denotes the part of the one-form along $\kappa$ (the vertical part)  and $\Omega^2_{\omega}$ denotes the part of the two-form along the symplectic form $\omega$. That is, a one-form and a horizontal, self-dual two-form can be written as
\begin{equation}
\begin{split}
\eta^{(1)}&=\kappaÊ\eta^{(0)}+\eta^{(1)}_H~, \\
\rho^{(2+)}_{H}&= \omega \rho^{(0)}+\rho^{(2,0)}_{H}+\rho^{(0,2)}_{H}~.
\end{split}
\end{equation}
Since the operator $L_{\phi}$ respects this decomposition, the determinant we are computing can be written as 
\begin{equation}\label{1loop1}
\begin{split}
h(\phi)&=\left(\frac{\text{det}_{\Omega^{(2,0)}_{H}}L_\sigma\cdot\text{det}_{\Omega^{(0,2)}_{H}}L_\sigma\cdot(\text{det}_{\Omega^{0}}L_\sigma)^3}{\text{det}_{\Omega_{H}^{1}}L_\sigma\cdot\text{det}_{\Omega^{0}}L_\sigma\cdot (\text{det}_{H^{0}} L_\sigma)^2}\right)^{\frac{1}{2}}  \\
&=\left(\frac{\text{det}_{\Omega^{0}}(iL_\sigma)\cdot\text{det}_{\Omega^{2,0}_{H}}(iL_\sigma)}{\text{det}_{\Omega_{H}^{1,0}}(iL_\sigma)}\right)^{\frac{1}{2}}\cdot\left(\frac{ \text{det}_{\Omega^{0}}(-iL_\sigma)\cdot \text{det}_{\Omega^{0,2}_{H}}(-iL_\sigma)}{\text{det}_{\Omega_{H}^{0,1}}(-iL_\sigma)}\right)^{\frac{1}{2}}\cdot \frac{1}{\text{det}_{H^{0}} L_\sigma}~.
\end{split}
\end{equation}   
Here we have inserted a factor of $i$, since the operator $iL_{\phi}$ has real eigenvalues. Since the eigenvalues can be both positive and negative, there is a possibility of a non-trivial phase of $h(\phi)$, and we write
\begin{equation}
h(\phi)=e^{-\frac{i\pi}{4}\eta}\left|h(\phi)\right|~.
\end{equation}
Here $\eta$ is defined by
\begin{equation}
\eta=\sum_{\lambda}{\text{sign}(\lambda)}~,
\end{equation}
where $\lambda$ are the eigenvaules of $iL_{\phi}$.
The factor $\eta$ needs regularization. In the standard way, we define
\begin{equation}\label{eta}
\eta(s)=\sum_{\lambda}{\text{sign}(\lambda)|\lambda|^{-s}}~.
\end{equation}
Below, we will carefully evaluate this expression at $s=0$, and find a $\phi$-dependence. It will therefore contribute non-trivially to the matrix model. 

To compute the above determinants, we need to decompose the spaces of horizontal differential forms into eigenspaces of the operator $\mathcal{L}_v$. Following \cite{Beasley:2005vf}, for a manifold $M$ which is a $U(1)$ fibration over a base manifold $\Sigma$, we can do this in the following way:
\begin{equation}\label{fourier}
\begin{split}
\Omega^{\bullet}_H(M,\mathfrak{g})&=\bigoplus_{t\neq 0}\Omega^{\bullet}(\Sigma,\mathcal{L}^t\otimes \mathfrak{g})\bigoplus\Omega^{\bullet}(\Sigma,\mathfrak{g})~.
\end{split}
\end{equation}
Here $\mathcal{L}$ denotes the line bundle associated to the $U(1)$ fibration. For each element $\xi_{t}\in\Omega^{\bullet}(\Sigma,\mathcal{L}^{t}\otimes\mathfrak{g})$, $\mathcal{L}_v$ acts as
\begin{equation}
\mathcal{L}_v\xi_t = 2\pi it\xi_t~.
\end{equation}
All fields, also the connection, transform in the adjoint under a gauge transformation with a constant parameter. We can decompose the Lie algebra into root spaces,
\begin{equation}
 \mathfrak{g}=\bigoplus_{\beta}\mathfrak{g}_{\beta}~,
 \end{equation}
$\beta$ denotes the roots. On $\mathfrak{g}_{\beta}$, $[\phi,~]$ has eigenvalue $i\langle \phi,\beta\rangle$, where $\langle~,~\rangle$ denotes the pairing between the Lie algebra and its dual. The eigenvalue of each mode $\xi_{t,\beta}\in\Omega^{\bullet}(\Sigma,\mathcal{L}^t\otimes\mathfrak{g}_\beta)$ is thus 
\begin{equation}\label{eigenvalue}
\lambda_{t,\beta} = 2\pi i t +i\langle \phi,\beta\rangle~, \quad t\in \mathbb{Z}~. 
\end{equation}
To determine the cancelation between the nominator and denominator in $\eqref{1loop1}$, we use the Atiyah-Singer index theorem. Inspection of $\eqref{1loop1}$ shows that the number of left over modes for each $t$ will be given by the index of the operators 
\begin{equation}
\begin{split}
&\bar{\partial}_V : \Omega^{(0,0)}(\Sigma,V) \rightarrow  \Omega^{(0,1)}(\Sigma,V)\rightarrow \Omega^{(0,2)}(\Sigma,V)~, \\
&\partial_V : \Omega^{(0,0)}(\Sigma,V)\rightarrow  \Omega^{(1,0)}(\Sigma,V)\rightarrow \Omega^{(2,0)}(\Sigma,V)~ ,
\end{split}
\end{equation}
that is, the Dolbeault complex twisted by the line bundle $V=\mathcal{L}^t$. The Atiyah-Singer theorem says that the indices of the operators $\bar{\partial}_V$ and $\partial_V$ are given by 
\begin{equation}\label{ivb}
\begin{split}
\text{ind}(\bar{\partial}_V)&=\int\limits_{\Sigma}{\text{Td}(TM^{+})\wedge \text{ch}(V)}~, \\
\text{ind}(\partial_V)&=(-1)^{\frac{n}{2}}\int\limits_{\Sigma}{\text{Td}(TM^{-})\wedge \text{ch}(V)}~,
\end{split}
\end{equation} 
where $n=\text{dim}\Sigma$, $TM^{\pm}$ denotes the holomorphic and anti-holomorphic tangent bundles, $\text{Td}(X)$ denotes the Todd class of the vector bundle $X$ and $\text{ch}(V)$ denotes the Chern character of the vector bundle $V$.  

We can write out $\text{Td}(TM^\pm)$ and $\text{ch}(V)$ in terms of Chern classes. We denote the $n$'th Chern class by $c_n$. We find
\begin{equation}
\begin{split}
\text{Td}(TM^+)&=1+\frac{1}{2}c_1(TM^{+})+\frac{1}{12}\left(c_1(TM^+)\wedge c_1(TM^+)+c_2(TM^+)\right)\ldots \\
\text{Td}(TM^-)&=(-1)^{\frac{n}{2}}\left(1-\frac{1}{2}c_1(TM^+)+\frac{1}{12}\left(c_1(TM^+)\wedge c_1(TM^+)+c_2(TM^+)\right) \ldots \right)\\
\text{ch}(V)&=1+c_1(V)+\frac{1}{2}c_1(V)\wedge c_1(V)+\ldots 
\end{split}
\end{equation}
where the last expression follows since $V$ is a line bundle in our case. We therefore find
\begin{equation}\label{index}
\begin{split}
\text{ind}(\bar{\partial}_V)&= \int_{\Sigma}{\left(\frac{1}{12}\left[c_1(TM^+)^2+c_2(TM^+)\right]+\frac{1}{2}c_1(TM^{+})\wedge c_1(V)+\frac{1}{2}c_1(V)^2\right)} \\
&=1+\frac{3}{2}t+\frac{1}{2}t^2 ~,\\
\text{ind}(\partial_V)&= \int_{\Sigma}{\left(\frac{1}{12}\left[c_1(TM^+)^2+c_2(TM^+)\right]-\frac{1}{2}c_1(TM^{+})\wedge c_1(V)+\frac{1}{2}c_1(V)^2\right)} \\
&=1-\frac{3}{2}t+\frac{1}{2}t^2 ~.
\end{split}
\end{equation}
In the last step we have used that on $\Sigma=\mathbb{C}P^2$, the total Chern class is written as $c(TM^+)=(1+\omega)^3$, where $\omega$ is the K\"ahler form normalized so that the integral of $\omega^2$ gives 1. We have also used that the first Chern class for the line bundle $V=\mathcal{L}^t$ is given by $c_1(\mathcal{L}^t)=tc_1(\mathcal{L})=t\omega$, since for line bundles, the first Chern class is multiplicative. For a nice review of the Atiyah-Singer index theorem and related matters, see for example \cite{Eguchi:1980jx}.
\subsection{Calculation of the absolute value of $h(\phi)$}
Combining $\eqref{1loop1},\eqref{fourier},\eqref{eigenvalue}$ and $\eqref{index}$, we find that the absolue value of $h(\phi)$ is given by

\begin{equation}\label{det23}
|h(\phi)|=\prod_{\beta}{\prod_{t\neq0}{\left(2\pi  t+\langle \phi,\beta\rangle\right)}\cdot \prod_{t\neq 0}{\left(2\pi  t+\langle \phi,\beta\rangle\right)}^{\frac{t^{2}}{2}}}~.
\end{equation}
We can evaluate the products over $t$ by using $\zeta$-function regularization. The Riemann zeta-function is defined by 
\begin{equation}
\zeta(s)=\sum_{t>0}{\frac{1}{t^s}}~.
\end{equation}
The sum is convergent for $\text{Re}s>1$, and is defined by analytic continuation for other values of $s$. 

The first term in $\eqref{det23}$ we can write as
\begin{equation}
\begin{split}
\prod_{\beta}{\prod_{t\neq0}{\left(2\pi  t+\langle \phi,\beta\rangle\right)}}=\prod_{t>0}{\left(2\pi t\right)^{2\Delta_G}}\cdot \prod_{\beta>0}\prod_{t>0}{\left(1-\frac{x^2}{t^2}\right)^2}~, \quad \quad x = \frac{\langle \beta,\phi\rangle}{2\pi}~,
\end{split}
\end{equation}
where $\Delta_G$ denotes the dimension of the $G$. Using the formula
\begin{equation}
\prod_{t>0}{\left(1-\frac{x^2}{t^2}\right)}=\frac{\sin(\pi x)}{\pi x}~, 
\end{equation}
and that 
\begin{equation}
\begin{split}
&\prod_{t>0}{2\pi}=e^{\ln{2\pi}\cdot \sum_{t>0}{1}}=e^{\ln{2\pi}\cdot \zeta(0)}=\left(2\pi\right)^{-\frac{1}{2}} ~,\\
&\prod_{t>0}{t}=e^{\sum_{t>0}{\ln{t}}}=e^{-\zeta'(0)}=\left(2\pi\right)^{\frac{1}{2}}~, \\
&\Rightarrow \\
&\prod_{t>0}{2\pi t}=1~,
\end{split}
\end{equation}
we evaluate the first term in $\eqref{det23}$ to 
\begin{equation}
\prod_{\beta>0}{\langle\beta,\phi\rangle^{-2}\left[2\sin\left(\frac{\langle\beta,\phi\rangle}{2}\right) \right]^{2}}~.
\end{equation}
The second factor in $\eqref{det23}$ we can write as
\begin{equation}
\prod_{t>0}{(2\pi t)^{\Delta_G t^2}}\cdot \prod_{\beta>0}\prod_{t>0}{\left(1-\frac{x^2}{t^2}\right)^{t^2}}~,\quad \quad x = \frac{\langle \beta,\phi\rangle}{2\pi}~.
\end{equation}
With the same method as above, the first factor evaluates to 
\begin{equation}
\prod_{t>0}{(2\pi t)^{\Delta_G t^2}}=e^{-\Delta_G \zeta'(-2)}~,
\end{equation}
where we have used the value $\zeta(-2)=0$. The second factor we can write as $\prod_{\beta>0}{e^{f(x)}}$, with 
\begin{equation}\label{f1}
f(x)=\sum_{t>0}{t^2\ln{\left(1-\frac{x^2}{t^2}\right)}}~.
\end{equation}
We evaluate this sum by taking the derivative with respect to $x$:
\begin{equation}
\begin{split}
f'(x)&=\sum_{t>0}{\left(-\frac{2x}{t^2}\right)\cdot \frac{t^2}{1-\frac{x^2}{t^2}}}=-2\left(\sum_{t>0}{\left(\frac{x(t^2-x^2)}{t^2-x^2}+\frac{x^3}{t^2-x^2}\right)}\right) \\
&=-2\sum_{t>0}{x}-2\sum_{t>0}{\frac{x^3}{t^2-x^2}} \\
&=-2\cdot \left(-\frac{x}{2}\right)-2\cdot x^3\cdot \frac{(1-\pi x \cot(\pi x ))}{2x^2} \\
&=\pi x^2 \cot(\pi x) \\
&\Rightarrow \\
&f(x)=\frac{i \pi x^3}{3}+x^2\ln(1-e^{-2\pi i x})+\frac{i x \text{Li}_2(e^{-2\pi i x})}{\pi}+\frac{ \text{Li}_3(e^{-2\pi i x})}{2\pi^2} + \text{constant}~.
\end{split}
\end{equation} 
We can fix the integration constant by demanding that $\eqref{f1}$ and the above expression gives the same Taylor expansion around $x=0$. That gives us $\text{constant}=-\frac{\zeta(3)}{2\pi^2}$. So we find
\begin{equation}
\begin{split}
&\prod_{\beta}{\prod_{t\neq 0}{\left(2\pi  t+\langle \phi,\beta\rangle\right)}^{\frac{t^{2}}{2}}}=e^{\Delta_G \zeta'(-2)}\cdot\prod_{\beta>0}{e^{f(x)}}~,\quad \quad x= \frac{\langle \beta,\phi\rangle}{2\pi} ~,\\
&f(x)=\frac{i \pi x^3}{3}+x^2\ln(1-e^{-2\pi i x})+\frac{i x \text{Li}_2(e^{-2\pi i x})}{\pi}+\frac{ \text{Li}_3(e^{-2\pi i x})}{2\pi^2} -\frac{\zeta(3)}{2\pi^2}~.
\end{split}
\end{equation}
We notice that $f(x)\rightarrow -\infty $ when $x\rightarrow \pm 1$, whereas for $-1<x<1$, $f(x)$ can be expanded as
\begin{equation}
f(x)=\sum_{k=0}^{\infty}{(-1)^k\frac{2^{2k}B_{2k}}{(2+2k)(2k)!} x^{(2+2k)}\pi^{2k}}~,
\end{equation}
where $B_{n}$ are the Bernoulli numbers. 

In summary, we have shown that the absolute value of the product of determinants in $\eqref{1loop1}$ is given by
\begin{equation}\label{abs}
\begin{split}
|h(\phi)|&=e^{-\Delta_G \zeta'(-2)}\cdot \prod_{\beta>0}{\left[\frac{\sin(\pi x)}{\pi x}\right]^2\cdot e^{f(x)}},\quad \quad x= \frac{\langle \beta,\phi\rangle}{2\pi} ~,\\
f(x)&=\frac{i \pi x^3}{3}+x^2\ln(1-e^{-2\pi i x})+\frac{i x \text{Li}_2(e^{-2\pi i x})}{\pi}+\frac{ \text{Li}_3(e^{-2\pi i x})}{2\pi^2} -\frac{\zeta(3)}{2\pi^2}~.
\end{split}
\end{equation}
 \subsection{The phase}
For $\eta$, we write
\begin{equation}
\eta=\eta_{+}(iL_{\phi})+\eta_{-}(-iL_\phi)~.
\end{equation}
Here, the first factor comes from the determinant on the space of holomorphic forms, and the second one from anti-holomorphic. Most terms in $\eta$ will cancel out, the Atiyah-Singer index theorem gives us the number of non-matching modes for each $t$. We find that the regularized expression $\eqref{eta}$ is given by
\begin{equation}
\begin{split}
\eta(s)=&\sum_{t,\beta}{\left( 1+\frac{3}{2}t+\frac{1}{2}t^2\right)\text{sign}(\lambda(t,\beta))|\lambda(t,\beta)|^{-s}} -\sum_{t,\beta}{\left(1-\frac{3}{2}t+\frac{1}{2}t^2\right)\text{sign}(\lambda(t,\beta))|\lambda(t,\beta)|^{-s}} \\
&=3\sum_{t,\beta}{t\cdot \text{sign}(\lambda(t,\beta))|\lambda(t,\beta)|^{-s}}~.
\end{split}
\end{equation}
We can rewrite this expression as
\begin{equation}
\begin{split}
&\sum_{t,\beta}{t \cdot \text{sign}(\lambda(t,\beta)|\lambda(t,\beta)|^{-s}}  \\
&=2\sum_{t>0,\beta>0}{\frac{t}{(2\pi t+\langle \beta,\phi \rangle)^{s}}}+2\sum_{t>0,\beta>0}{\frac{t}{(2\pi t-\langle \beta,\phi \rangle)^{s}}}+2\Delta_T \sum_{t>0}{\frac{t}{(2\pi t)^s}}~.
\end{split}
\end{equation}
Here, $\Delta_T$ is the dimension of the Cartan subalgebra for $G$, and we have without loss assumed that $0<\frac{\langle \beta,\phi \rangle}{2\pi }<1$. 

Expanding this expression for small $\langle \beta,\phi \rangle$, we find 
\begin{equation}
\begin{split}
&\frac{4}{(2\pi)^s}\sum_{t>0,\beta>0}{t^{-(s-1)}}+2\Delta_T \sum_{t>0}{\frac{t}{(2\pi t)^s}}+\frac{2}{(2\pi)^{s+2}}\sum_{t>0,\beta>0}{s(s+1)\cdot t^{-(s+1)}\langle \beta,\phi \rangle^2} +\mathcal{O}(s) \\
&=-\frac{2}{12}\Delta_G+\frac{1}{2\pi^2}\sum_{\beta>0}{\langle \beta,\phi \rangle^2} \\
&=-\frac{1}{6}\Delta_G-\check{c}_\mathfrak{g}\frac{1}{2\pi^2}\text{Tr}\phi^{2}~.
\end{split}
\end{equation}
Above, we have identified the Riemann $\zeta$-function, and in the middle step used its value at $s=-1$ and its expansion for small $s$ around $s+1$. $\check{c}_\mathfrak{g}$ denotes the dual Coexter number of the group $G$, and we have used the formula $\sum_{\beta>0}{\langle \beta,\phi\rangle^2}=-\check{c}_\mathfrak{g}\text{Tr}\phi^{2}$. $\Delta_G$ denotes the dimension of $G$. Since the higher order terms in the first line brings extra factors of $t$ in the denominator, the sum of $t$ will converge and evaluating at $s=0$ gives zero. Therefore, we will only get a $\text{Tr}\phi^{2}$ correction to the matrix model from the phase of the determinant. So finally, we find the phase of the determinant in $\eqref{1loop1}$ to be given by the expression
\begin{equation}\label{phase}
\eta(0)=-\frac{1}{2}\Delta_G-\check{c}_\mathfrak{g}\frac{3}{2\pi^2}\text{Tr}\phi^{2}~.
\end{equation}

In conclusion, combining the above expressions $\eqref{abs}$ and $\eqref{phase}$, we find $h(\phi)$ to be
\begin{equation}
\begin{split}
h(\phi)&=e^{-\frac{i\pi}{4}\eta(0)}\cdot |h(\phi)| \\
&=e^{\frac{i\pi}{8}\Delta_G+\check{c}_\mathfrak{g}\frac{3i}{8\pi}\text{Tr}\phi^{2}}\cdot e^{-\Delta_G \zeta'(-2)}\cdot \prod_{\beta>0}{\left[\frac{\sin(\pi x)}{\pi x}\right]^2\cdot e^{f(x)}}~, \quad \quad x = \frac{\langle \beta,\phi\rangle}{2\pi} ~,
\end{split}
\end{equation}
where
\begin{equation}
f(x)=\frac{i \pi x^3}{3}+x^2\ln(1-e^{-2\pi i x})+\frac{i x \text{Li}_2(e^{-2\pi i x})}{\pi}+\frac{ \text{Li}_3(e^{-2\pi i x})}{2\pi^2} -\frac{\zeta(3)}{2\pi^2}~.
\end{equation}

\bibliographystyle{utphys}
\providecommand{\href}[2]{#2}\begingroup\raggedright\endgroup

\end{document}